# Innovative SiC ultraviolet instrumentation development with potential applications for the Habitable Worlds Observatory

**Prabal Saxena,[a,*] Zeynep Dilli,[b] Peter Snapp,[a] Tilak Hewagama,[a] Shahid Aslam,[a] Chullhee Cho,[a] Augustyn Waczynski,[a] Nader Abuhassan,[c] Anh T. La,[a] Bryan K. Place,[c] Allison Youngblood,[a] Thomas F. Hanisco,[a] Ryan Stauffer,[a] Dina Bower,[a] Akin Akturk,[b] Neil Goldsman,[b] Bryce Galey,[b] Ethan Mountfort,[b] Mitchell Gross,[b] Ryan Purcell,[b] Usama Khalid,[b] Yekta Kamali,[b] Chris Darmody,[b] Robert Washington,[d] Tim Livengood,[a,e,f] Daniel P. Moriarty,[a,e,f] Roser Juanola-Parramon,[a,g] Carl A. Kotecki,[a] Narasimha S. Prasad,[h] and Joseph Wilkins[d]**

[a]NASA Goddard Space Flight Center, Greenbelt, Maryland, United States
[b]CoolCAD Electronics, Inc., Greenbelt, Maryland, United States
[c]SciGlob Instruments and Services, LLC, Columbia, Maryland, United States
[d]Howard University, Washington, District of Columbia, United States
[e]University of Maryland, Department of Astronomy, College Park, Maryland, United States
[f]Center for Research and Exploration in Space Science and Technology II, Greenbelt, Maryland, United States
[g]University of Maryland, Baltimore County, Baltimore, Maryland, United States
[h]NASA Langley Research Center, Hampton, Virginia, United States

**ABSTRACT.** We detail recent and current work that is being carried out to fabricate and advance novel SiC UV instrumentation that is aimed at enabling more sensitive measurements across numerous disciplines, with a short discussion of the promise such detectors may hold for the Habitable Worlds Observatory. We discuss SiC instrument development progress that is being carried out under multiple NASA grants, including several PICASSO and SBIR grants, as well as an ECI grant. Testing of pixel design, properties, and layout as well as maturation of the integration scheme developed through these efforts provides key technology and engineering advancement for potential HWO detectors. Achieving desired noise characteristics, responsivity, and validating the operation of SiC detectors using standard readout techniques offers a compelling platform for the operation of denser and higher dimensionality SiC photodiode arrays of interest for use in potential HWO coronagraph, spectrograph, and high-resolution imaging instruments. We incorporate these SiC detector properties into a simulation of potential NUV exoplanet observations by HWO and also discuss potential applications to HWO.

## 1 Introduction

The most recent Astrophysics Decadal Survey Report[1] underscored the importance of ultraviolet (UV) observations in exploring numerous high-priority science questions. UV observations are key to understanding stellar atmospheres, the morphology and dynamics of galaxies, the interaction of black holes with their surroundings, and the properties of worlds outside our solar system, including their potential habitability. This is merely a subset of areas where UV observations are key, and such observations were also recognized as an essential means of achieving the objectives of the Decadal Report's top flagship priority, the Habitable Worlds Observatory (HWO). The driving exoplanet and cosmological objectives of HWO require observations in the UV range of 100 to 300 nm using normal incidence optics.[2] These HWO will require UV detectors with sensitivity and noise properties significantly improved over the current state-of-the-art (SOA). In this paper, we describe ongoing work to advance novel silicon carbide (SiC) detectors that would enable more sensitive UV measurements across numerous disciplines through their ability to achieve potentially photon-noise limited imaging at room temperature (theoretical $3.5 \times 10^{-18}$ $e^{-}/s/\text{pixel}$ thermally generated dark current for 10 $\mu$m square pixels at 270 K) and insensitivity to contaminant visible light (QE $< 7 \times 10^{-3}\%$ at wavelengths beyond 400 nm), with a short discussion of the promise such detectors may hold for HWO.

*Address all correspondence to Prabal Saxena, prabal.saxena@nasa.gov

The near-UV (NUV) spectral region (defined in this work as ~200 to 300 nm) is a subset of the UV spectral region that contains many powerful tracers of atoms, molecules, and small particulates such as dust and hazes. These NUV tracers have wide-ranging applications across numerous disciplines, such as Astrophysics, Planetary Science, Heliophysics, and Earth Science. Here, we briefly describe science cases where SiC detectors would provide substantial gains over SOA silicon-based focal plane arrays.[3] A key advantage in all cases is SiC's diminished sensitivity to visible light, which can often swamp desired UV signals. With respect to Planetary Science, order-of-magnitude signal-to-noise ratio improvements in UV spectroscopy of comet comae, planetary atmospheres, and aurorae would address key planetary decadal objectives.[4] SiC detectors can also enable new measurements of molecules on bodies that are episodically sourced from plumes, for example, measurements of airglow emission when Io is in shadow (which could not historically be observed, due to the long accumulation time and contamination from scattered light from Jupiter), and could reveal the relative effects of $SO_2$ frost and $SO_2$ gas photoabsorption.[5] Furthermore, improved SNR is critical for observing terrestrial bodies under challenging solar illumination conditions, such as observations of polar regions on the Moon. This is timely, as NASA's Artemis and CLPS programs, as well as other international agencies, have targeted the lunar south polar region for human and robotic surface exploration within the next several years. Improved NUV observations of the lunar south pole have the ability to provide insight into latitude-dependent space weathering processes and probe compositional diversity and volatile distribution across the region at traverse-relevant spatial resolution, both providing context and guiding site selection and surface science activity prioritization.[6–8] With respect to Heliophysics, SiC detectors have the potential to serve as radiation particle detectors in solar observations that require blindness to interfering visible and longer wavelength UV photons.[9] Applications specific to Earth Science include ground- or space-based passive remote sensing of trace gases and aerosols. Optical light filtering is required for SOA Si detectors[10] and can interfere with accurate chemical retrievals by adding fringe structures to collected spectra that can obscure weaker chemical signatures. Although SiC detectors have numerous other applications in each of these disciplines, this article focuses on technology development and application to an HWO astrophysics use case.

With respect to Astrophysics, SiC detectors have the potential to provide a wide range of sensitive, enabling observations of interest, including UV spectroscopy of active galactic nuclei, exoplanet host stars, and gravitational wave sources. SiC detectors directly satisfy the need for high-quantum efficiency (QE), solar-blind, broad-band NUV detectors identified in the Astrophysics Biennial Technology Report 2022.[11] In fact, the highest priority science areas from the Astrophysics Decadal Survey Report[1] all benefit from sensitive NUV observations. For example, the NUV is a critical bandpass for distinguishing between physical models[12] of the extremely hot and rapidly evolving kilonova that emerges in the seconds to hours after the merger of two neutron stars. In addition, in the search for habitable exoplanets, the Hartley-Huggins ozone band in the NUV is the most sensitive tracer of oxygenic atmospheres.[13] Obtaining information regarding these atmospheres is complicated by the high contrast ratios between the host star and planet and the low amount of reflected light from potential exo-Earth candidates. To enable significant breakthroughs in these areas, next-generation detectors must maximize in-band sensitivity while minimizing noise. For imaging applications, minimizing responsivity to longer wavelength light is essential because many astrophysical sources are orders of magnitude brighter in the optical than in the NUV. SiC possesses many of these properties[3] and, with the advancement of fabrication techniques, is now being explored by our team as a potential option for many of these use cases.

We are pursuing fabrication of SiC instruments relevant to numerous disciplines under multiple projects funded by NASA's Science Mission Directorate (SMD) and Space Technology Mission Directorate (STMD), as well as through Small Business Innovation Research Grants and NASA Goddard Space Flight Center's Internal Research and Development program. The PReSSiC (Planetary Remote Sensing using SiC detectors - PIs Saxena/Hewagama) project[14] is building, testing, and validating a high-sensitivity silicon carbide focal plane detector optimized for miniaturized instrumentation that can be part of remote sensing planetary science missions, including cubesats. The focal plane assembly would obtain radiance observations with a higher detectivity than previous generations of NUV spectrometers from 200 to 340 nm at resolving power $R \sim 200$, enabling SNR $> 10\times$ better than current leading detectors from identical observational platforms. The key science motivation for PReSSiC is understanding the hydration state of planetary bodies across the solar system, with a specific focus on the moon and the identification of outgassed (non-water-related) volatiles, characterization of atmospheric constituents and hazes, and understanding of the mineralogy of airless bodies. Simultaneously, an ongoing development effort is the NASA STMD Early Career Initiative-supported "Design of a Broad Utility, Visible-Blind, SiC, UV Spectrometer for Astrophysics, Planetary and Earth Science Applications, Validated Through Representative Measurement of Climate Forcing Gases" project (PI Snapp), hereafter referred to as the Pandora+SiC project. Pandora+SiC aims to achieve clean differential optical absorption spectroscopy of columnar trace gases in Earth's atmosphere in the 280 to 450 nm range through integration of visible-blind UV detectors, produced by leveraging recent advances in SiC fabrication, into an existing, ground-based, atmospheric spectroscopy platform, Pandora.[15] SiC instrument development is also being carried out under another NASA ROSES PICASSO award, namely, the Ultraviolet Detector Innovation for Raman Exploration and Characterization (UV-DIRECT) of Ocean Worlds project (PI Bower). UV-DIRECT aims to enable identification of minerals, volatiles, organic molecules, biopolymers, water, and other hydrous phases in planetary materials by fabricating a SiC avalanche photodiode (APD), with high internal gain, specifically targeted for applications to ocean worlds' surface missions that require robust and compound-specific instrumentation. UV-DIRECT aims to enable identification of minerals, volatiles, organic molecules, biopolymers, water, and other hydrous phases in planetary materials by fabricating a SiC APD, with high internal gain, specifically targeted for the identification of ocean world-relevant compounds.[16] UV-DIRECT is being carried out in collaboration with Army Research Labs and is developing a detector that would use UV/NUV (266 to 340 nm) Raman spectroscopy with ppb sensitivity to achieve these planetary science observations.[17] Raman spectroscopy is mainly an *in situ* technique, but because SiC APD technology is so versatile, the development efforts and results for UV-DIRECT have fed into the other projects more relevant to HWO.[18] Although a range of SiC instrumentation development is being carried out by our team, we will focus on the PReSSiC and Pandora+SiC plans and development in this paper and how advancements in these projects may be beneficial for future potential Astrophysics applications.

Section 4 describes key background information on SiC as a semiconductor material, with a brief discussion of comparison to other materials. Section 5 lays out planned and current development work in the key SiC instrument development projects that are in progress. Section 6 provides updates on the fabrication and integration progress of the photodetectors for the corresponding projects. Section 7 discusses results and updates on recent and current testing of photodetector properties as well as planned future work. Finally, Sec. 8 incorporates these SiC detector properties into a simulation of exo-Earth observations by HWO.

## 2 Background on SiC

Work on the development of highly sensitive wide energy band gap ($E_g$) UV-EUV detectors at NASA Goddard Space Flight Center (GSFC) started more than two decades ago,[3] particularly in direct band gap material AlGaN ($E_g = 3.4 - 6.2$ eV)[19–21] and indirect band gap material 4H-SiC ($E_g = 3.2$ eV)[22,23] material systems with successful fabrication of photodiodes demonstrating good performance characteristics.[3,24]

For NASA, these wide energy band gap devices offered several key advantages for space applications over conventional Si ($E_g = 1.1$ eV)-based devices. These include visible-blind detection, high thermal stability, better radiation hardness, high breakdown electric field, high chemical inertness, and greater mechanical strength. Additional detail on these properties is given later in this section. Furthermore, the shorter cutoff wavelength of these material systems eliminates the need for bulky and expensive optical filtering components, mitigating risk and allowing for simpler optical designs, and most importantly, increasing in-band throughput.

The most compelling reason for using wide-band gap semiconductors for UV detectors is the low dark currents that can be attained with no or minimal cooling, significantly reducing the technical complexity needed to achieve photon-noise–limited imaging. The dark current, under zero or reverse bias conditions in the depletion region of a semiconductor, is determined from the rate of generation ($G$) of electron-hole pairs

$$G = \frac{n_i}{\tau},$$

where $n_i$ is the intrinsic carrier concentration and $\tau$ is the effective recombination lifetime. The intrinsic carrier concentration is given by

$$n_i = \sqrt{N_v N_c \exp\left(-\frac{E_g}{2kT}\right)},$$

where $N_v$ is the valence band density of states, $N_c$ is the conduction band density of states, $E_g$ is the energy band gap, $k$ is the Boltzmann constant, and $T$ is the temperature. The dark current density, $J_d$, is directly proportional to $n_i$ and is given by

$$J_d = \frac{q n_i W}{\tau},$$

where $q$ is the electronic charge and $W$ is the depletion width. For silicon at room temperature, $n_i$ halves for every 5.5 K decrease in temperature.[25] By comparison, a wider band gap semiconductor with the same density of states and $\tau$ as silicon, but a 0.036 eV higher band gap, achieves the same factor of two reduction in $n_i$. The exponential dependence of $n_i$ on band gap makes a 0.12 eV increase in band gap equivalent to a 30 K temperature drop, both offering an order-of-magnitude reduction in thermal generation rate. Wide-band gap semiconductors achieve significantly lower thermally generated dark current due to orders of magnitude reduction in $n_i$, even in the presence of large differences in $\tau$, and therefore have applications in high temperature space environments (see Table 1). It can be seen that 4H-SiC at 300 K shows ∼18 to 19 orders of magnitude lower dark current than Si and that 4H-SiC has ∼3 orders of magnitude lower dark current than 6H-SiC due to the wider bandgap. At high temperatures, e.g., 737 K (nominal Venus surface temperature), 4H-SiC still maintains ∼10 orders of magnitude advantage over Si. The relationship between intrinsic carrier concentration versus temperature is shown in Fig. 1.

**Table 1** Comparison of dark current density generation in Si, 4H-SiC, and 6H-SiC.

| Parameter | Si | 4H-SiC | 6H-SiC |
|---|---|---|---|
| Bandgap (eV)[26–28] | 1.12 | 3.26 | 3.03 |
| Conduction band density of states, $N_c$ ($cm^{-3}$)[26–28] | $2.8 \times 10^{19}$ | $1.7 \times 10^{19}$ | $1.5 \times 10^{19}$ |
| Valence band density of states, $N_v$ ($cm^{-3}$)[26–28] | $1.0 \times 10^{19}$ | $2.5 \times 10^{19}$ | $5.0 \times 10^{18}$ |
| Intrinsic carrier concentration, $n_i$ ($cm^{-3}$) @ 300 K | $1.0 \times 10^{10}$ | $8.2 \times 10^{-9}$ | $1.6 \times 10^{-6}$ |
| Dark current density, $A/cm^2$, @ 300 K[a] | $8.0 \times 10^{-10}$ | $6.6 \times 10^{-28}$ | $1.3 \times 10^{-25}$ |
| Dark current density, $A/cm^2$, @ 737 K (Venus)[a] | $3.2 \times 10^{-1}$ | $4.5 \times 10^{-11}$ | $7.8 \times 10^{-11}$ |
| Normalized dark current density, $A/cm^2$, @ 300 K[a] | 1 | $8.2 \times 10^{-19}$ | $1.6 \times 10^{-16}$ |

[a]Assuming identical geometry, i.e., depletion width, $W = 10\ \mu$m and generation lifetime, $\tau = 1\ \mu$s.

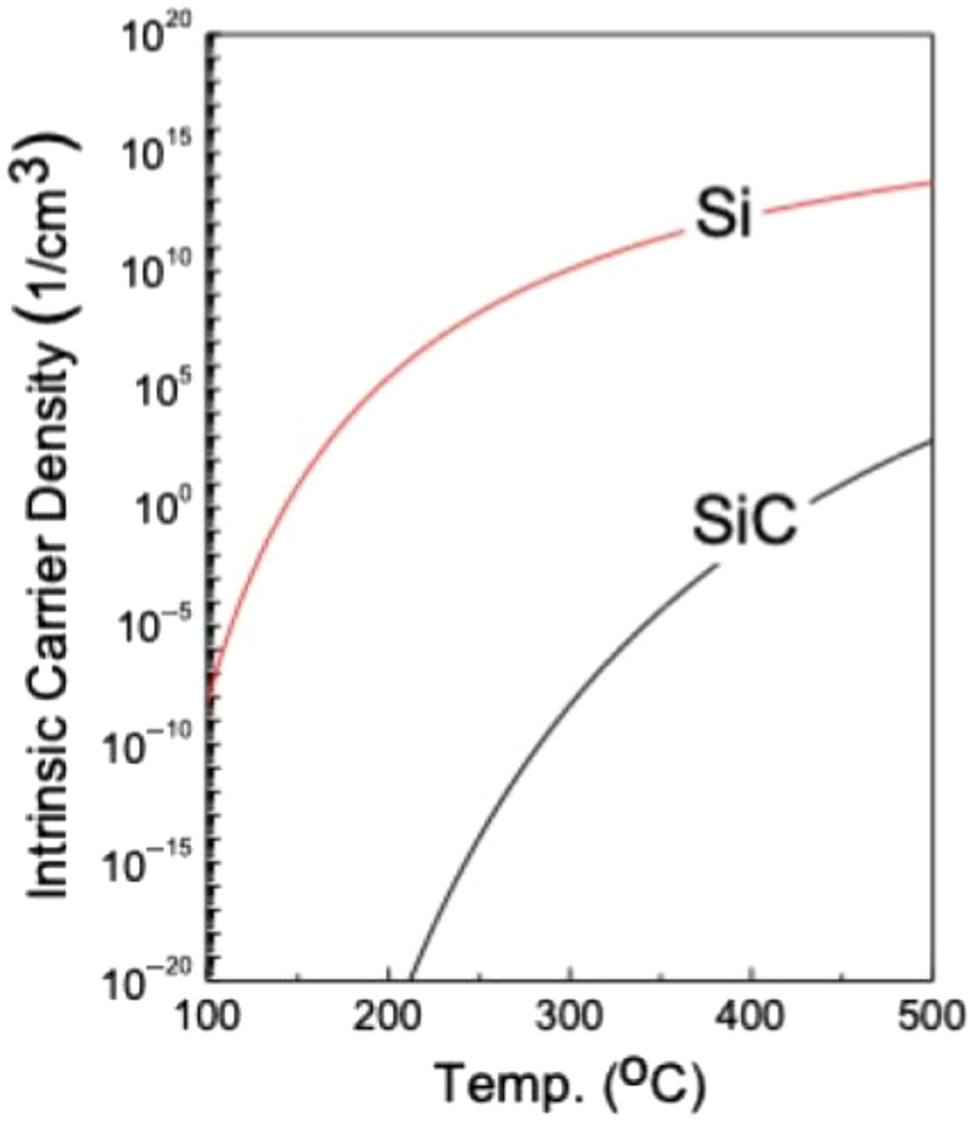

**Fig. 1** Comparison of intrinsic carrier density as a function of temperature for Si and SiC. At 270 K, the intrinsic carrier density for 4H-SiC is $7 \times 10^{-11}/\text{cm}^3$, which gives a theoretical thermally generated current density of $5.6 \times 10^{-30}$ $\text{A/cm}^2$. For a 10 $\mu$m square pixel at 270 K, this is equivalent to $3.5 \times 10^{-18}$ $e^-$/s/pixel (c.f. $1 \times 10^3$ $e^-$/s/pixel for Si).

Furthermore, wide-band gap semiconductors exhibit strong temperature dependence, indicating that moderate cooling can significantly reduce the dark current (see Fig. 1 and literature[9,29]). Dark current typically scales proportionally with pixel area, but this relationship is more nuanced than simple linear scaling. The primary thermal generation current component does scale directly with area as it originates from bulk generation-recombination processes throughout the depletion region volume. However, real photodetectors exhibit additional dark current components including surface leakage currents that scale with pixel perimeter rather than area, and defect-assisted generation currents that may concentrate at interfaces or crystallographic defects. Consequently, as pixel dimensions decrease, the area-to-perimeter ratio changes, often leading to higher dark current densities in smaller pixels due to the increased relative contribution of edge effects. This non-ideal scaling becomes particularly significant in deep submicron pixels, where surface states at pixel boundaries can dominate the dark current generation. Wide bandgap materials such as SiC benefit from dramatically reduced bulk generation currents, making them less sensitive to area scaling, but they remain subject to surface and interface effects. For precision coronagraph applications as needed for HWO that require long integration times, understanding this scaling relationship becomes critical for predicting detector performance across different pixel geometries and optimizing device design to minimize dark current contributions.

Wide-band gap semiconductors have additional benefits for detecting UV-A (315 to 400 nm), UV-B (280 to 315 nm), UV-C (200 to 280 nm), and EUV (1 to 125 nm) radiation because they are inherently visible blind. This visible blindness significantly reduces the need for blocking filters, which simplifies optical system design and reduces the size, weight, power consumption, and cost (SWaP-C) of UV instruments, and most importantly, increases the system QE, allowing for more sensitive measurements.

Compared with narrow-band gap materials, wide-band gap semiconductors are significantly more radiation tolerant, making them excellent candidates for space applications.[30–33] SiC offers superior radiation hardness (long-term resilience) compared with conventional Si due to its stable crystal structure resulting from stronger covalent bonds than Si, and thereby higher displacement threshold with low defect mobility, which reduces radiation-induced damage. SiC's stronger covalent bonds and higher melting point make it more resilient to radiation exposure, as it can maintain its structural integrity even under extreme conditions such as the high-radiation environment of space. When SiC gets cooled to lower temperatures, SiC's radiation resistance can be further enhanced owing to the reduced thermal vibrations and lower defect mobility preventing defects from migrating or clustering. Moreover, the reduction in charge carrier mobility at lower temperatures prevents radiation-induced disruptions in SiC's electrical performance, making it an ideal candidate for space applications.

In summary, SiC detectors have several advantages over Si detectors:

1. Wide bandgap: The band gap of 4H-SiC is 3.2 eV, three times wider than Si. Its intrinsic carrier density is $10^{20}$ times lower than that of Si, leading to significantly lower dark current.
2. Visible blindness: SiC detectors are visible-blind, enabling UV, EUV, and X-ray photon detection without interference from visible or IR backgrounds.
3. Excellent radiation hardness: SiC has high displacement energy (~21 eV) compared with Si (12 eV), offering superior radiation hardness.

Although all wide-bandgap semiconductors such as AlGaN provide the advantages conveyed above in their ideal form, SiC has several practical advantages over similar wide-bandgap semiconductors, including material maturity and processability. For example, the defect density of SiC ($10 - 10^3/\mathrm{cm}^3$) is several orders of magnitude lower than GaN ($10^6 - 10^{10}/\mathrm{cm}^3$) and approaches that of Si ($0.5/\mathrm{cm}^3$), resulting in significantly better dark current performance when deployed in actual detectors. Similar materials proposed for UV detection, including indium gallium nitride (InGaN),[34] diamond,[35] and zinc oxide (ZnO),[36] suffer from a similar lack of material maturity and need complex processing to realize functional devices, whereas SiC can be procured in wafer form and processed similarly to Si. With respect to another similar material, SiC may be able to provide similar or improved measurements compared with current SOA delta-doped silicon CCDs utilized on missions such as SPARCS[37] without a need for additional filters [see sample QE curve for an SiC diode in Fig. 7(a)]. Much of this is because SiC substrate and epi-growth technologies have developed to such a level as to allow the fabrication of many different types of SiC photodetectors with desired features.[9] SiC UV p-i-n photodiodes have already been fabricated and are commercially available. SiC APDs with extremely high gain and low excess noise have also been demonstrated, including single photon avalanche diodes (SPAD), which demonstrate gains ($1.4 \times 10^6$ at 116.8 V reverse bias) comparable to microchannel plates (MCP) used for UV astronomy ($10^6 - 10^7$) but in completely solid-state packages and at much lower operating voltages (few hundred volts for SPAD versus few kilovolts for MCP). Zhou et al.[38] demonstrated $1 \times 128$ linear arrays of 4H-SiC UV APDs with high pixel yield (100%), uniform breakdown voltage with a variation of 0.2 V, QE of 53.5% at 285 nm, and dark currents below 1 nA (100 $\mu$m diameter circular detector, $6.24 \times 10^9$ $e^-/\mathrm{sec}/\mathrm{pix}$) at 95% of breakdown voltage. For additional information on the progress and history of SiC fabrication and use in many other applications, see Ref. 9.

## 3 Overview of Development Activities

SiC instrument development by our team under the current programs has leveraged the expertise of scientists, engineers, and commercial companies with fabrication, integration, and implementation experience. Key to the fabrication tasks across all the projects is the work that CoolCAD Electronics, LLC, has been carrying out using unique expertise in the design and fabrication of devices and integrated circuits with SiC. CoolCAD has previously worked on SiC low-power CMOS, deep-UV sensors, and building blocks for deep-UV-sensitive focal plane arrays as well as high-power, high-voltage, and high-temperature SiC devices. CoolCAD's current work to produce UV sensors and readout circuitry on the same substrate leverages background work supported by NASA SBIR programs (NNX17CG32P, "Silicon Carbide 10 $\mu$m-Pitch UV Imaging Array and APD with Active Pixel Readout," Phase 1 SBIR; NNX16CG51P, "Single Chip EUV, VUV, and Deep UV Photodetector System with Integrated Amplifier," Phase 1 SBIR; NNX17CG15C, "A Silicon Carbide Foundry for NASA's UV and High Temperature CMOS Electronics Needs," Phase 2/2E SBIR). CoolCAD has demonstrated two classes of 4H-SiC UV sensor devices and structures, both of which are top-lit: PN-junction and Schottky diodes, both as single diodes and in focal plane arrays, which are the focus of the current projects. The spectral responsivity of a given SiC diode is, among other factors, dependent on the location and depth of the photocurrent-collecting junction electric field. Schottky diodes, with the built-in field closest to the surface, are expected to show a higher responsivity at shorter wavelengths, which has been observed in fabricated devices. In PN junction devices, which are produced via implanting dopants at specific levels and concentrations in a SiC lattice, CoolCAD employs implant profile engineering to set the implant energies and doses necessary to achieve a desired profile, which has also been incorporated in current work. Finally, both past GOES-R funding and a current IRAD are supporting detailed demonstration of the radiation hardness and visible blind nature of SiC photodiodes, which has begun to be elucidated in the literature (maintenance of responsivity up to fluences of $10^{13}$, 600 keV, $He^+$ ions [90 kGy]), through analysis in space-relevant radiation environments, 63 MeV protons ($7 \times 10^{10}$ to $2.5 \times 10^{12}$ $cm^{-2}$) to mimic 5 years of cumulative on-orbit exposure development.

The PReSSiC project aims to deliver a lab-tested $1 \times 128$ pixel 4H-SiC detector suitable for spectrometers in miniaturized platforms that can obtain radiance observations with a higher detectivity than previous generations of NUV spectrometers from 200 to 340 nm at $R \sim 200$. Key development for this project has been testing a number of design trades and optimizing design and fabrication for this goal. The project has begun with a focus on the design and layout of individual sensors and small arrays. Potential single-pixel geometries and fabrication parameters are being evaluated to optimize device characteristics such as responsivity and response speed, which will be examined in the context of the science requirements. A key area of exploration is a test of implant depth, as it is valuable to create an implant profile that supports enhanced sensitivity in the spectral region of interest. There are limits on this process, such as the difficulty of getting implants past a certain depth into the tough SiC crystal material. An additional consideration is that in a focal plane array, the pixels must be electrically and optically isolated by deep trenches for clear image formation. CoolCAD Electronics has previously demonstrated small-pixel-count 2D arrays with 10 $\mu$m pitch that had not featured deep trench isolation, but is examining deep isolation trenches in the new arrays.

PReSSiC is fabricating a number of different test structures and small-pixel-count arrays to identify potential fabrication difficulties and limits to our fabrication techniques. The Pandora +SiC program fabrication run (middle [$1 \times 128$ focal plane arrays + process control modules], right [$1 \times 128$ and $1 \times 1024$ focal plane arrays + process control modules]). The dies on both Pandora+SiC program wafers also have variation in pixel pitch (25 and 50 $\mu$m), pixel height (200 and 500 $\mu$m), and implant density (N−/P and N+/P junction diodes). Diode pitch is a key parameter for FPA performance and is being varied between 6 and 40 $\mu$m to establish factors limiting diode miniaturization. In addition to pitch, PReSSiC is looking to optimize other factors influencing SiC photodiode performance including contact resistance to minimize current losses at interfaces, minimum trench width (including the overall necessity of isolation trenches) to balance pixel isolation (and therefore clear image formation) with array density, patterning long metal lines for effective electrical interconnection lines, surface leakage reduction to improve noise performance, and identification of additional factors affecting responsivity and dark current. The team has also started examining packaging and input/output (I/O) structure design for both active and passive pixels. The team is further designing and building a set of readout electronics, particularly for the planned scaling up of the pixel count for the full array. Radiometric testing is being conducted and will be conducted at numerous steps along the way using a testbed that is currently under assembly so that the arrays can eventually be tested under both ambient and vacuum conditions. Key performance testing related to responsivity, noise performance, and dispersion will be compared with requirements based upon simulations.

Another key ongoing development effort is the NASA Early Career Initiative-supported "Design of a Broad Utility, Visible-Blind, SiC, UV Spectrometer for Astrophysics, Planetary and Earth Science Applications, Validated Through Representative Measurement of Climate Forcing Gases" project, hereafter referred to as the Pandora+SiC effort. The Pandora+SiC effort aims to develop and integrate visible-blind UV detectors into an existing, ground-based, atmospheric spectroscopy platform, Pandora,[15] which has been developed by a commercial partner, SciGlob, LLC. Pandora+SiC will improve the reliability and temporal resolution of total column formaldehyde measurements conducted by Pandora either for direct data collection or validation for data collected from satellite platforms such as the ozone monitoring instrument.[39] A simultaneous goal of this project is also to simplify instrument setup through the elimination of filter components, enabled by the visible blindness of SiC, and reducing or eliminating cooling requirements, enabled by the reduced dark current of SiC. Pandora+SiC is at the point of integrating fabricated SiC detectors, $1 \times 128$ and $1 \times 1024$ arrays of high aspect ratio ($25 \times 200$ $\mu$m, $25 \times 500$ $\mu$m, and $50 \times 500$ $\mu$m), passive, P-on-N photodiodes with external capacitive transimpedance amplifier (CTIA) read-out integrated circuits (ROICs) configured to be controlled by an external imaging array controller. Following integration, detector units will be tested to confirm QE greater than 0.1 at 345 nm and dark current <20 pA (individual diodes probed so far show a dark current upper limit of <6 fA [$750 \times 750$ $\mu$m square detector, $3.74 \times 10^4$ $e^-/\sec/\mathrm{pix}$] discussed in Sec. 5) and integrated into the Pandora spectrometer.

The current detector and electronics configuration for both the PReSSiC detectors and the Pandora+SiC detectors has been tailored for easy integration into planetary science instruments and Pandora instrument systems, respectively. Although not directly applicable to HWO, testing of pixel design, properties, and layout in the PReSSiC project and maturation of the integration scheme developed through the Pandora+SiC project both provide key technology and engineering advancement for potential HWO detectors. Achieving desired noise characteristics, responsivity, and validating the operation of SiC detectors using standard readout techniques offers a compelling platform for the operation of denser and higher dimensionality SiC photodiode arrays of interest for use in potential coronagraph, spectrograph, and high-resolution imaging instruments.

## 4 Fabrication and Integration Progress of Photodetectors

For the PReSSiC and Pandora+SiC projects, many of the initial tasks are focused on design and fabrication, and CoolCAD is working on a variety of linear photodiode structures in order to map the design space, identify design best practices, and optimize fabrication techniques.

Diode geometry is the first aspect being explored. For the particular trace gas spectroscopy application targeted by the Pandora+SiC project, the pixel pitch required for the target spectral resolution is 25 $\mu$m. The optical design of the instrument can take advantage of pixels with a high aspect ratio. Therefore, the initial fabrication run of the program includes pixels with $25 \times 500$ $\mu$m and $25 \times 200$ $\mu$m drawn areas (yielding sub 1 pA dark currents under test, $6.24 \times 10^6$ $e^-/\sec/\mathrm{pix}$), along with test structures with a different aspect ratio. Figure 2 shows micrographs from the fabrication process of these designs, defining different diode regions. In the context of the PReSSiC program, a major program target is miniaturization to a very small pixel pitch, and the devices being fabricated for this program include diodes with pitches as small as 10 and 6 $\mu$m. Follow-on work will focus on scaling production methods to cover large arrays (up to 8k × 8k) via transition to stepper lithography production methods and integration with dense 2D readout circuitry. Figure 3 shows fabrication micrographs of some of these miniaturized structures, including dopant implantation sites on a wafer which features diode arrays with 6, 10, 15, and 25 $\mu$m pitches. Other geometric features being explored are the placement and sizes of contacts and interconnects, to evaluate certain trade-off aspects such as fill factor versus diode series resistance.

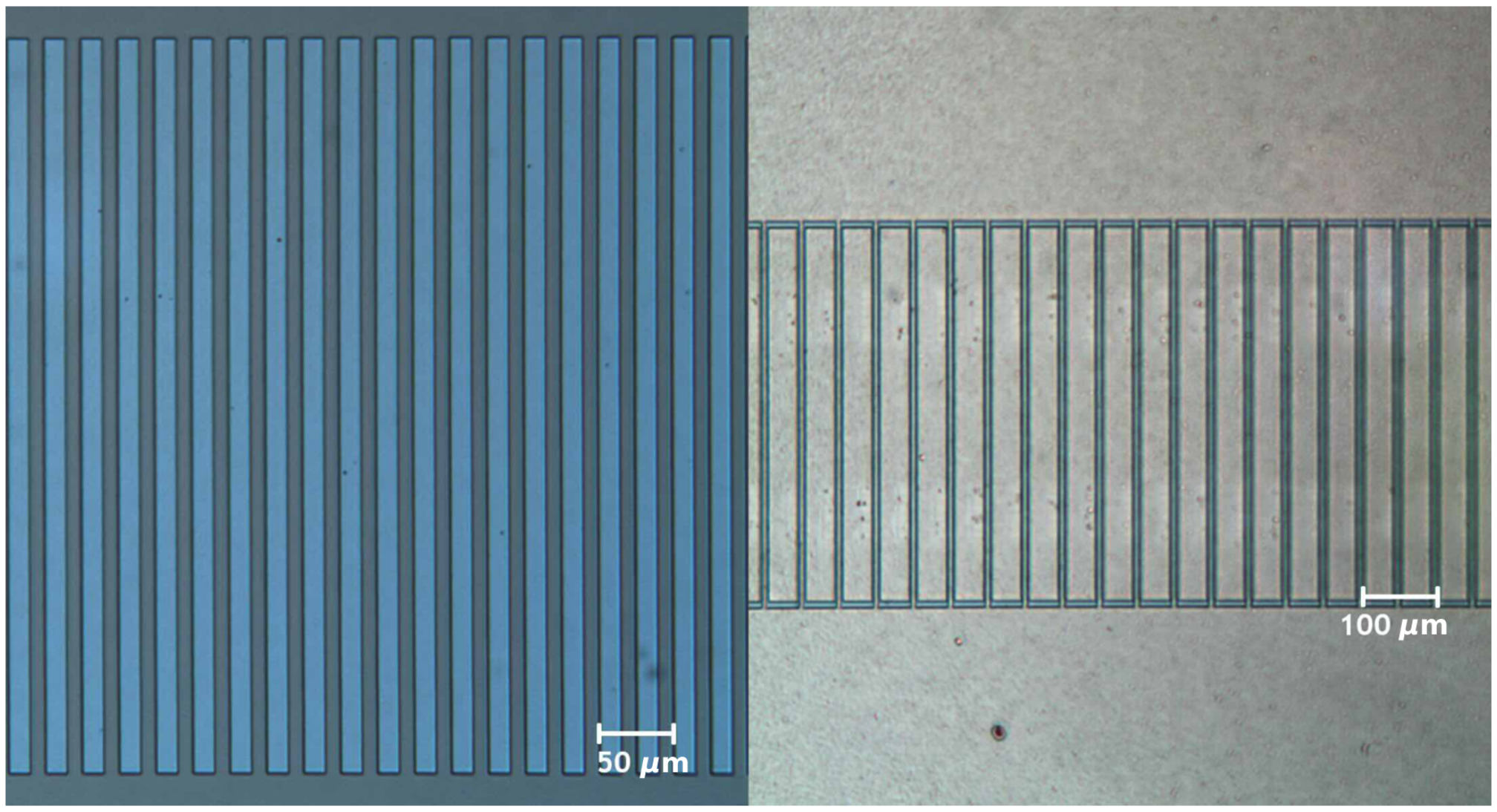

**Fig. 2** Micrographs from the process of different diode regions defined for subsequent ion implantation of the tall, narrow-aspect-ratio diodes being developed for spectroscopy applications for the Pandora+SiC project.

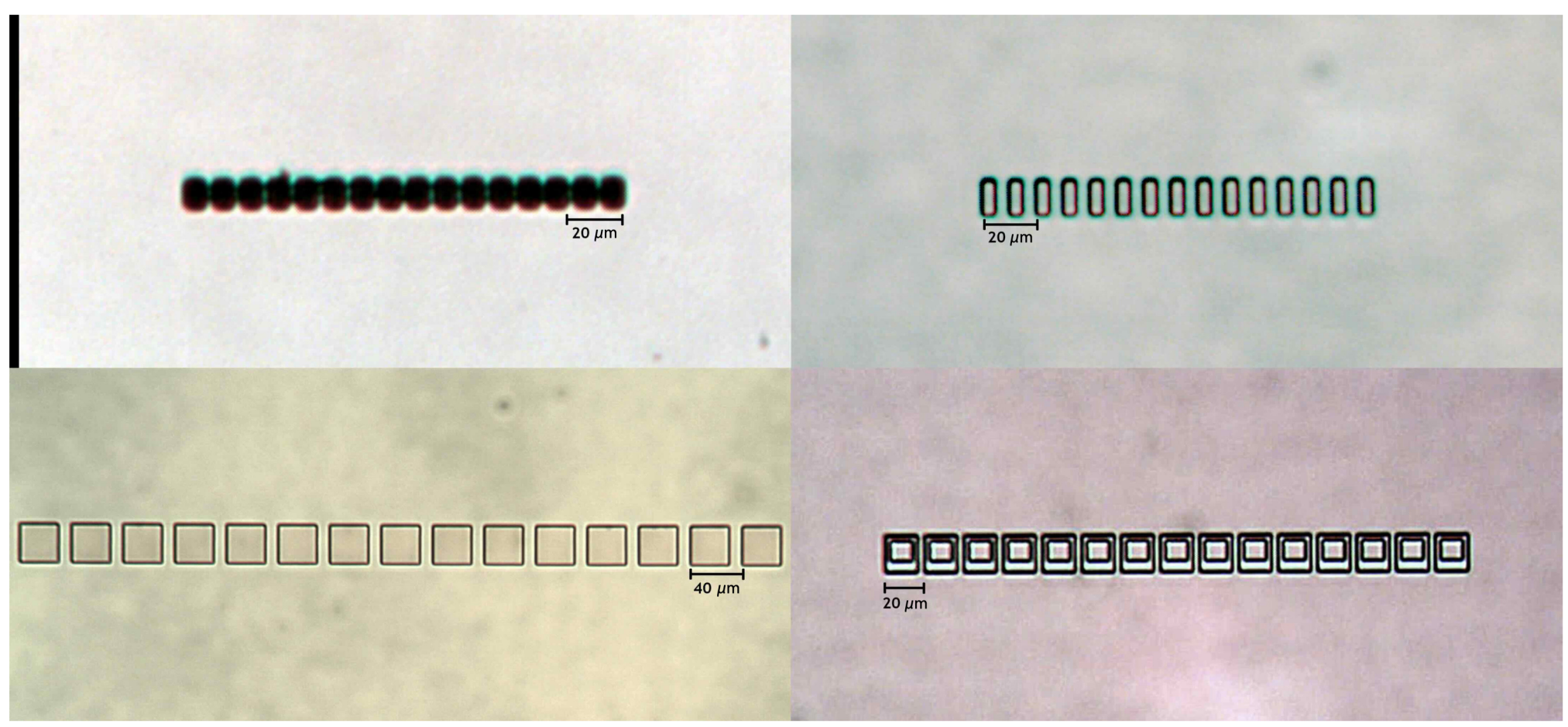

**Fig. 3** Micrographs from the process of different diode regions being defined by ion implantation for diode arrays with different pixel pitches for the PReSSiC project, targeting diode miniaturization.

Another aspect under exploration is the utility of different junction structures. Although shorter wavelengths have larger absorption coefficients and smaller absorption depths, the different junction structures, which define the position and depth of the depletion region where photogenerated charge carriers are collected, are expected to have somewhat different spectral responsivity. To focus on broadband responsivity pushing toward 350 nm, the PReSSiC and Pandora+SiC programs are fabricating N−/P and N+/P junction diodes with different doping depths and concentrations, yielding different geometries and depths for the photocurrent collection regions. This approach shifts both the peak wavelength response as well as the bandwidth of the total response, allowing tuning for specific measurements. To optimize response at shorter wavelengths for a separate NASA SBIR program, we are also fabricating Schottky diodes, which have the photocurrent collection region starting directly at the upper surface of the diode, where shorter wavelengths are readily absorbed. Planned further developments include the use of different epitaxial structures, pre-dopant implantation, or metallization for Schottky junction formation, to allow for more sophisticated diode fabrication as well.

Beyond individual pixel size and responsivity, in applications that use diode arrays, such as spectroscopy or imaging, sufficient pixel-to-pixel isolation is required to allow clean data collection. For the Pandora+SiC program, deep trenches between the photoactive regions of each diode are being implemented for this purpose. For the PReSSiC program, in addition to trenches, we are also implementing array variations without trenches to evaluate ways to achieve pixel-to-pixel isolation with alternative diode geometries. This alternative approach allows us to increase the fill factor, especially of the smallest diodes. The development and evaluation of these isolation structures are being pursued in the SBIR program as well.

Figure 4 shows the wafer-scale designs for the initial runs in the PReSSiC project (left panel) and the Pandora+SiC program (right panel). The PReSSiC program design includes small-pixel-count arrays (15 × 1 and 8 × 1) and single test diodes. The Pandora+SiC program design includes 128 × 1 and 1024 × 1 arrays and single test diodes. Both designs are currently in fabrication and have progressed through their dopant implant and isolation trench formation steps. We are currently optimizing back-end processing steps, which comprise the metallization steps for contact metal and interconnect formation, preceding implementation of these steps on the PReSSiC and Pandora+SiC wafers to complete their fabrication.

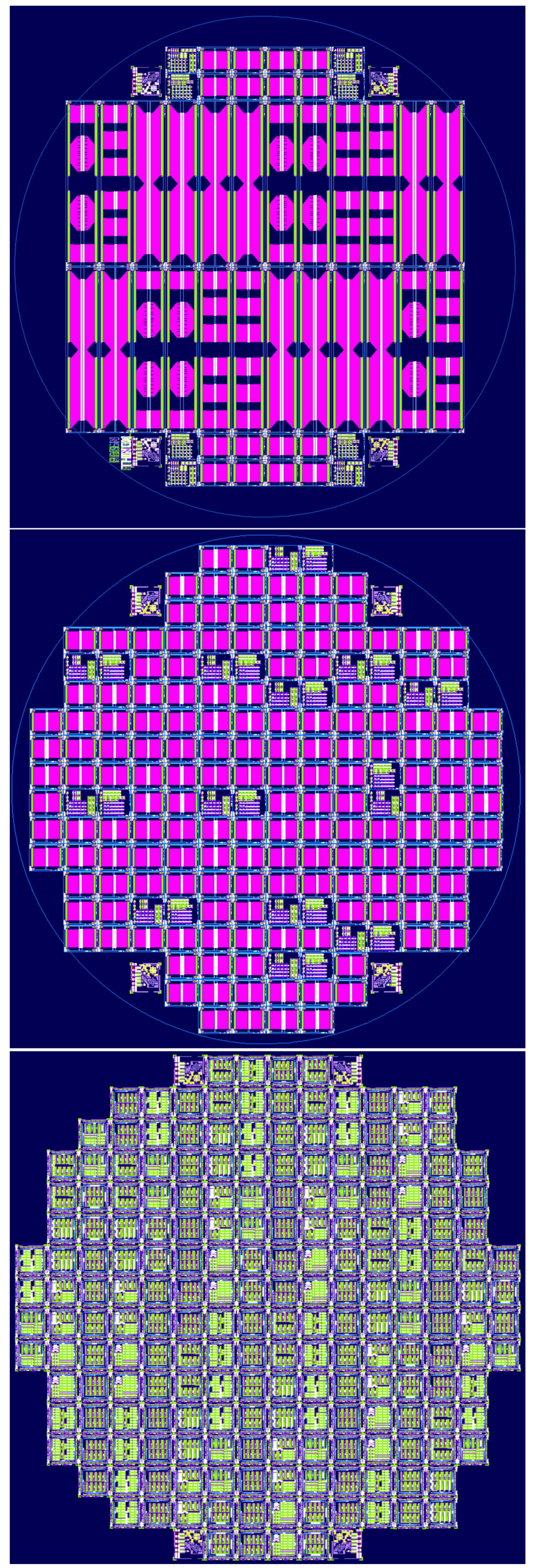

**Fig. 4** Wafer-scale layouts for the PReSSiC project fabrication run (left) and the Pandora+SiC program fabrication run (middle, right). The PReSSiC run design contains test structures and small-pixel count arrays. The middle design contains $128 \times 1$ arrays with some process control monitoring dies included. The design on the right contains both $1024 \times 1$ and $128 \times 1$ arrays and PCM dies. A more detailed description is given in the text.

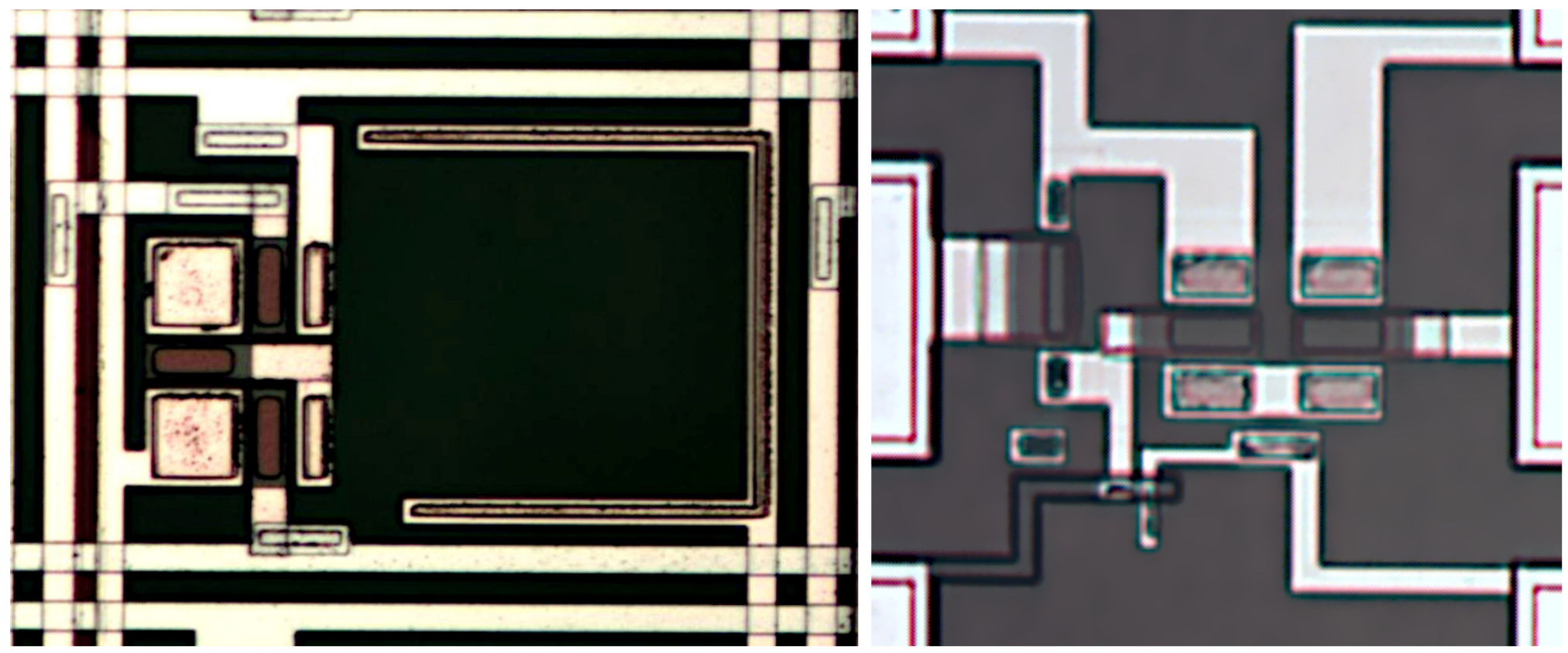

**Fig. 5** Previously fabricated photodiodes integrated with the first stage of a readout circuit on the same SiC die.

For both projects, the next stage of development will include active diode arrays, for which the first stage of a readout circuit converting the photocurrent to a voltage level over an integration period is fabricated on the same die as the photodiode arrays. CoolCAD Electronics has active development programs on SiC MOSFET structures and circuits and has previously demonstrated single diodes integrated with MOSFET-based readout circuits. Figure 5 shows previously fabricated examples of this structure. These development programs all start from the level of physics-based semiconductor device modeling and simulations and include process simulation and development efforts as well as device and circuit design, layout, and fabrication. For the PReSSiC project, in addition to one-dimensional arrays for spectroscopy, two-dimensional arrays with expanded pixel counts will also be implemented for push-broom spectroscopy and imaging applications. The pixel count will be expanded as well. In the SBIR program, CoolCAD is further developing larger-scale two-dimensional arrays, both of passive and active pixels. For the two-dimensional active array development, this effort includes work to miniaturize the transistors so they can be implemented within a given pixel, as opposed to being connected externally with metal fan-out, permitting effective scaling in two dimensions.

In addition to advancing fundamental understanding of SiC photodiode fabrication and performance, the Pandora+SiC effort is focused on a short-term field demonstration of SiC focal plane arrays in an existing atmospheric spectroscopy system, Pandora, for formaldehyde detection. As a result, the selection of diode array formats was informed by the need for a mature and highly reliable diode design, the ability to interface with external readout circuitry without complex hybridization techniques, and operability at low bias levels.

This all needs to be achieved while still allowing for modification of diode structure and pitch to achieve the necessary diode height to interact with the full spill of light from the Pandora grating, to afford a high enough spectral resolution for accurate capture of formaldehyde transmission signals. To that end, mature 1D arrays of PN photodiodes with electrical contacts on their upper surface that can fan out to wire bond pads matched to external readout electronics were selected [Fig. 6(a)].

P/N+ and P/N− implant variations are being evaluated to balance depletion region depth, and therefore responsivity, with serial resistance for optimized performance. Within these constraints, diode arrays with lateral pitches of 25 $\mu$m [Fig. 6(b)(i)] and 50 $\mu$m [Fig. 6(b)(ii)] and heights of 200 to 500 $\mu$m are being fabricated to best balance the total active area ($\sim$75% active area for 25 $\mu$m pitch devices due to the surface metallization cover, $\sim$85% active area for 50 $\mu$m pitch devices) and spatial resolution during final applications in the field.

To speed delivery, the Pandora+SiC effort has relied on fully passive photodiode arrays with each diode fanning out to a wirebond output pad, which can then be connected to the wirebond input pad of a corresponding ROIC. In this case, a set of four commercial CTIA CMOS readouts with wire bondable inputs is used [Fig. 6(a)]. ROICs will be situated on a basic PCB with output-boosting operational amplifiers to allow operation by an external commercial imaging array controller.

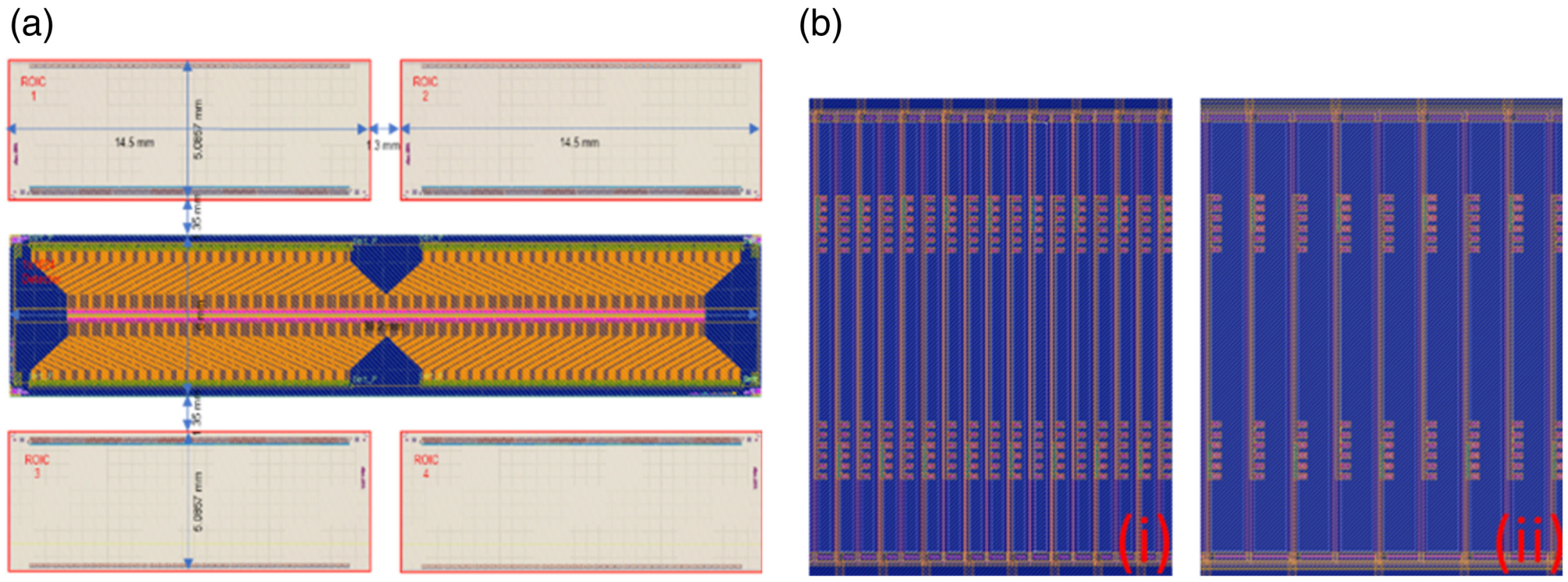

**Fig. 6** Structure of SiC photodiode-based focal plane arrays for the Pandora+SiC effort consists of (a) a core $1 \times 1024$ array of SiC photodiodes, which fans out to wire bond pads matched to the inputs of wire bondable ROICs. In this case, the array is interfaced with 4 CTIA ROICs, each with 256 inputs. (b) The Pandora+SiC effort is currently in the process of characterizing multiple diode variants, including (i) arrays with a 25 $\mu$m pitch, and (ii) arrays with a 50 $\mu$m pitch.

Although this is a simple approach to readout, it is an ideal starting point for developing readouts for more complex SiC arrays, as SiC can be read with biases consistent with those already matured for Si detectors ($\lesssim$2 V) with sufficient responsivity to fill detector wells in a reasonable exposure time. This includes generalization to reading out 2D SiC arrays with similar top-side fanout structures, as will be tested in the PReSSiC effort.

This approach can be further generalized to readout using commercial 2D ROICs with the maturation of top-illuminated SiC detector arrays with through-chip vias for electrical integration,[40] or backside-illuminated SiC detectors[41] electrically integrated using flip-chip, indium bump bonding techniques.

## 5 Current Testing of Photodetector Properties and Near-Term Future Work

Our team has conducted initial characterization and testing of diodes during the development and fabrication process. Of note is that because much of the work is still in early stages, these photodiode characteristics are not yet optimized, and in some cases, measurements are limited by evaluation techniques or hardware. Where this is the case, we provide a minimum possible performance parameter but expect SiC photodiodes to significantly exceed values reported here. This section describes some of the tests that have been conducted already and notes their limitations where applicable.

To achieve high-quality formaldehyde trace signature retrieval using the Pandora platform for the Pandora+SiC effort, individual photodiodes in the arrays described previously must meet minimum performance requirements. In particular, a minimum QE of 0.14 at 300 nm is necessary to accumulate sufficient signal within a reasonable acquisition time for the selected ROIC, and dark current must be below 20 pA to achieve an appropriate signal-to-noise ratio.

Preliminary testing on isolated large photodiodes ($\approx$0.5 mm$^2$) with implant structures similar to those targeted for the Pandora+SiC effort (with different profiles) meet or exceed these requirements, achieving a QE of 0.1 at 345 nm, with QE peaking at $\approx 0.4 - 0.5$ at 280 nm [Fig. 7(a), measurements performed at 0 V bias]. Performance is even more promising in terms of dark current. Extracting dark noise from a long-duration measurement of photodiode current under dark conditions and in the absence of any cooling or additional stabilization [Fig. 7(b), data collected at 0 V bias] establishes a ceiling on dark current of $\sim$6 fA ($750 \times 750$ $\mu$m square detector, $3.74 \times 10^4$ $e^-/\mathrm{sec}/\mathrm{pix}$, less than 7 $e^-/\mathrm{sec}/\mathrm{pix}$ for a $10 \times 10$ $\mu$m pixel). We claim this as a ceiling as this value is likely a result of shot noise arising from the transresistance amplifier used for the measurement. This performance should conservatively allow for photon noise-limited measurement (for an 11-ms integration would collect $1.6 \times 10^{-6}$ signal electrons while dark current represents only 687 electrons, and readnoise represents only 500 electrons) with signal-to-noise ratios in excess of 1172 during measurement.

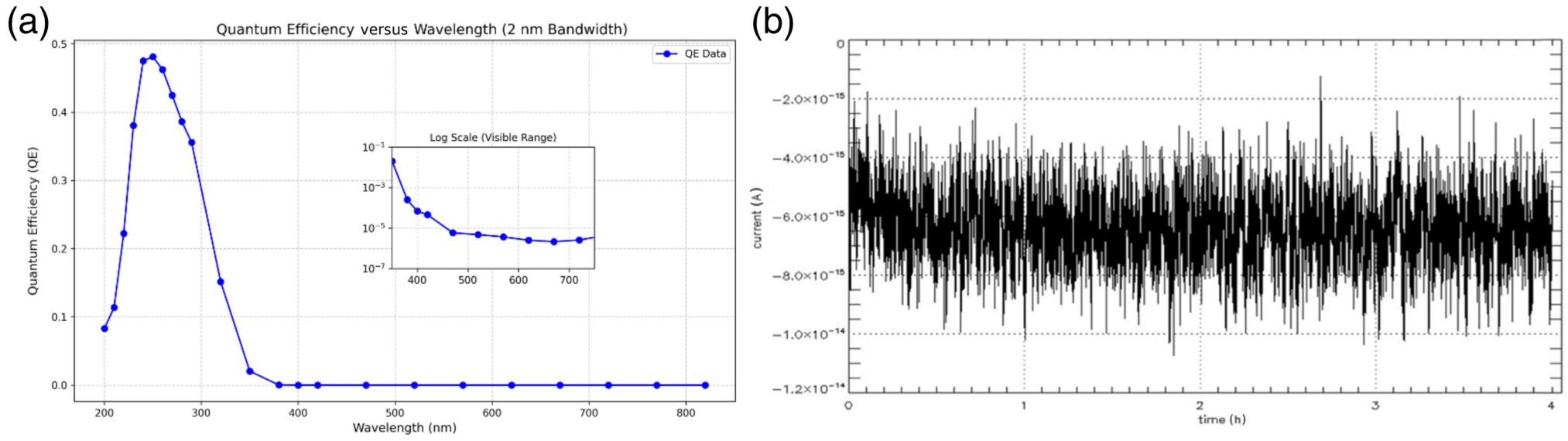

**Fig. 7** Characterization of large format SiC photodiodes ($750 \times 750$ $\mu$m square) for Pandora+SiC effort including (a) broad spectrum QE characterization and (b) dark current measurement. Note, although no bias is applied during measurement, the diode was connected in a forward bias configuration (positive terminal on P-side, negative terminal on N-side), and dark current flows from N to P (this manifests as a negative current during measurement). Additional development work is expected to improve QE values, although current values already satisfy Pandora+SiC goals, whereas dark current is an upper bound on the true value given measurement limitations.

Moving forward, analysis of dark current and QE will be expanded by probing detectors coupled to the selected CTIA ROIC with integration capability and correlated double sampling, allowing for more accurate quantification of dark current by removing the effects of shot noise. This step is essential for accurately assessing the utility of SiC photodiodes in astrophysics applications.

Similarly, initial testing of the I-V curves and QE has been conducted for three different photodiodes designed under an SBIR program that is a precursor to the PReSSiC program. The first is an $n^+$-$p$ diode, the second an $n^-$-$p$ diode, and the third, a Schottky diode. The $n^+$-$p$ and $n^-$-$p$ diodes are both 40-$\mu$m pitch diodes with exposed active areas of about $172.5 \times 10^{-6}$ mm$^2$. The Schottky diode is also a 40-$\mu$m pitch diode with an exposed active area of $270 \times 10^{-6}$ mm$^2$. I-V curves were constructed by measuring current values, using transresistance amplifiers without integration, under illumination from 200 to 350 nm, in 5 nm increments, at varying bias values. QE values were also calculated for all three diodes in the same wavelength range and are shown in Fig. 8.

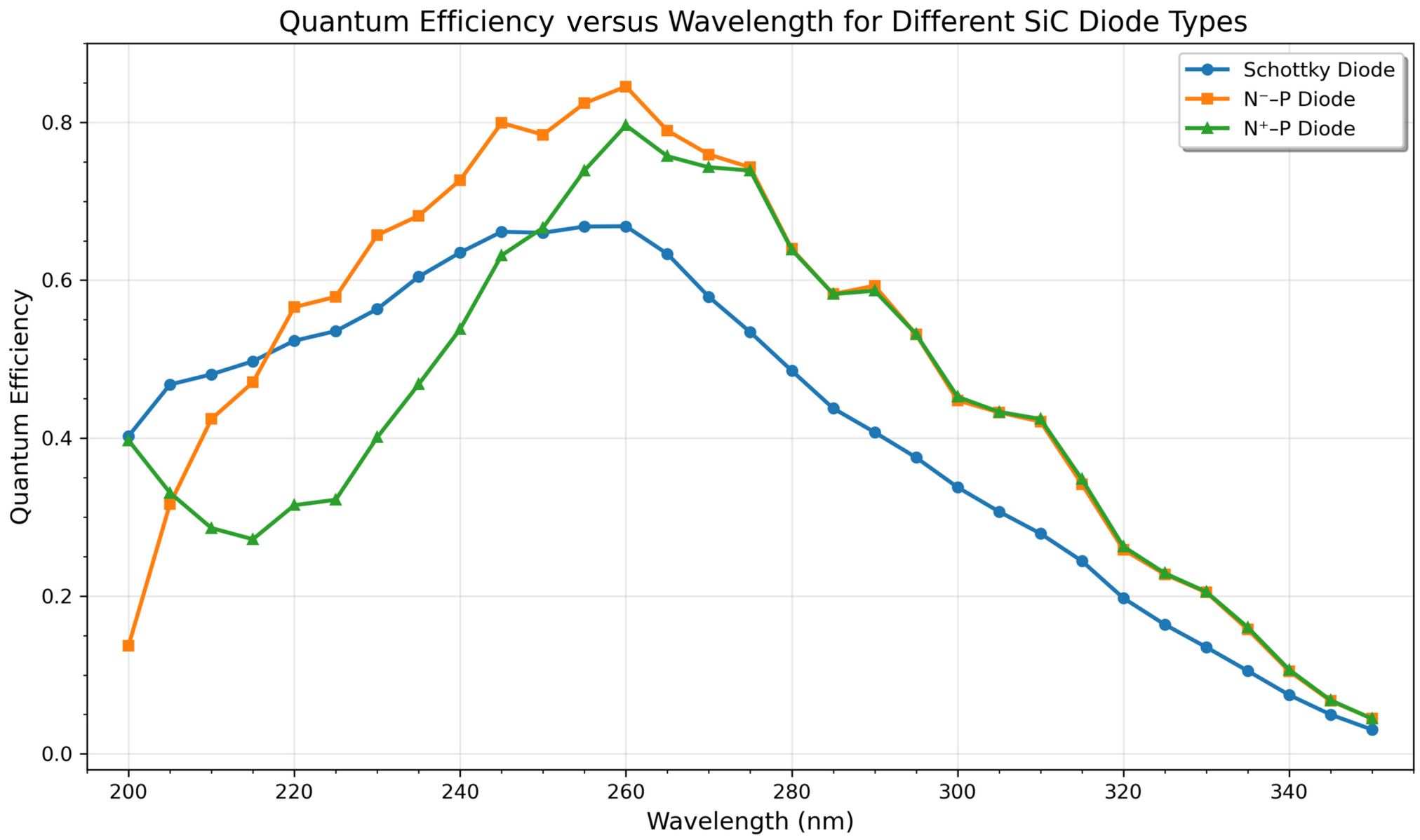

**Fig. 8** QE versus wavelength for different diode types produced for an SBIR program that is a precursor to the PReSSiC program. The plot compares the spectral response of three diode configurations: Schottky, N$^-$-P, and N$^+$-P diodes across the 200 to 350 nm wavelength range.

**Fig. 9** Comparison of NEP and specific detectivity of SiC diodes.

QE values for the $n^+$-$p$ diode peaked at 0.8 at 260 nm, were above 0.6 from 240 to 280 nm, and were above 0.2 from 200 to 330 nm. QE values for the $n^-$-$p$ diode peaked at a little bit above 0.8 at 260 nm, were nearly at or above 0.6 from 230 to 290 nm, and were above 0.2 for a similar wavelength range as the $n^+$-$p$ diode. QE values for the Schottky diode peaked from 250 to 260 nm at nearly 0.7, were 0.6 or higher from 230 to 270 nm, and were above 0.2 from 200 to 320 nm. Optical I-V curves exhibited a similar shape and general trend across all three diodes, although we note there was an offset in voltage where current approached zero, which was likely due to an instrumental effect. Using the dynamic resistance calculated from the I-V curves and the QE values, we were able to calculate specific detectivity ($D^*$) and noise equivalent power (NEP) values for each of the diodes over the 200- to 350-nm wavelength range.

The values for NEP and $D^*$ for the diodes are given in Figs. 9(a) and 9(b). The three diodes exhibit broadly the same relationships and trends versus wavelength, with the $n^+$-$p$ and $n^-$-$p$ diodes yielding slightly lower NEP values. The 230 to 300 nm range possesses the most favorable noise characteristics across all three diodes, with a drop-off in sensitivity toward longer wavelengths. Although these measurements are important initial characterization steps for the development of these detectors, we note that they are minimum bounds on the sensitivity of the diodes. This is because measurements of the I-V curve were done at coarse spectral resolution, and more highly sampled measurements where current drops to zero are likely to yield even lower dynamic resistance values for the diodes. In addition to the measurement improvements in both Pandora+SiC and the SBIR/PReSSiC characterization efforts, we note that in both cases, these are early-stage diodes in the development process. We expect that as design and fabrication processes mature, NEP and $D^*$ values will improve.

Future characterization and testing include improvements to both testing methodology and hardware as well as additional tests that correspond to future development needs. Enabling very low current measurements, including signal integration for correlated double sampling, and carrying measurements out at highly sampled voltage values near a current minima is a high near-term priority. In the future, as additional diodes from upcoming production runs are available for testing, we will conduct additional tests to optimize diode parameters prior to focal plane array/spectrometer design. Once that is achieved, and a focal plane array and spectrometer are fabricated, our team will utilize pre-existing optical testbeds for the Pandora+SiC and PReSSiC projects to carry out instrument testing.

## 6 Promising Detector Material for the Habitable Worlds Observatory

The HWO was recommended by the most recent astrophysics Decadal Survey[1] to support broad coverage of 100 nm – 2.5 $\mu$m wavelength through a variety of instrument modes, including high-contrast direct imaging of circumstellar material and bodies through coronagraphy, high angular resolution imaging, integral field spectroscopy, multi-object spectroscopy, and single-object spectroscopy. NUV capability of each of these modes is required to advance our understanding of exoplanet atmospheres, stars, interstellar medium, galaxies, and solar system science. At this early stage in the mission concept design, the HWO Technology Maturation Project Office has adopted instruments from the LUVOIR-B mission concept[42] as a starting design and is additionally considering a UV coronagraph instrument (G. Arney, private communication).

There are no requirements yet for these instruments, but based on the LUVOIR report, there is a clear need for large-format arrays that are buttable on three sides and have high QE, low dark current, and low read noise. Some instrument modes may not require large-format arrays but may have more stringent requirements on QE and noise. The LUVOIR telescope was planned to be actively controlled to an operating temperature of 270 K and had set limits on mass and power for different observatory segments.[43] SiC detectors have the potential to satisfy or achieve many of these requirements and desired properties, with theoretical 4H-SiC photodiodes predicted to have as little as $3.5 \times 10^{-18}$ $e^-/s/pixel$ thermally generated dark current when realized as 10 $\mu m$ square pixels, which would considerably outperform the EMCCDs ($1.4 \times 10^{-6}$ $e^-/s/pixel$ for 13 $\mu m$ resolving element size) identified as meeting the performance goals in the identified 2024 Astrophysics Technology Biennial Report's Technology Gap List Descriptions. In the rest of this section, we focus on a UV coronagraphy application for SiC.

The key scientific motivation for the UV coronagraph is to search for ozone on potentially habitable exoplanets. Ozone is a significant biosignature on the Earth because it is generated from the photolysis of oxygen by UV sunlight. The Hartley-Huggins band of ozone between 200 and 350 nm is much stronger than the Chappuis band near 600 nm and would be a sensitive tracer of ozone at levels significantly lower than present-day values. This is key because oxygen and consequently ozone are likely to have varied significantly over Earth's history, and over much of the Earth's past are likely to have been at much lower levels than present-day values. For example, prior to the great oxygenation event around 1 to 2 billion years ago, the abundance of oxygen in Earth's atmosphere was likely to have been only 0.1% of present-day levels.[44] Current oxygen and ozone levels are a relatively recent feature of Earth's atmosphere, and consequently, there is increasing recognition that sensitivity to smaller abundances of the two molecules is desirable when probing whether a world may be habitable. Thus, NUV spectroscopy is a powerful probe of smaller amounts of ozone that may enable characterization of a planet's habitability even when key biosignature gases are at abundances significantly lower than present-day Earth's abundances.

To demonstrate the promise of SiC detectors for HWO, we describe a use case of measuring ozone in an Earth-like exoplanet with the UV coronagraph in comparison to other potential detector materials such as Si and AlGaN. We assume the UV coronagraph would be operable over the 225 to 450 nm range, but with a smaller instantaneous bandpass and optimized for 250 nm. We conduct simulations over the 210 to 350 nm range to capture the Hartley-Huggins ozone band and align with wavelength regions of focus for the two SiC development projects detailed in previous sections. Future SiC development projects could optimize for different wavelength regions. We assume we are observing a potential exo-Earth candidate at a distance from Earth that corresponds to the Tau Ceti system (3.7 pc). Multiple HWO target lists indicate that such a system would have favorable observational properties for HWO,[45] including accessibility of the system's entire habitable zone[46] (although we note there are likely even more favorable systems that are closer to Earth).

We assume atmospheric profiles for the simulated planet from Whole Atmosphere Community Climate Model—WACCM6 runs that simulate atmospheres relevant to Early Earth.[47] We specifically choose oxygen and corresponding ozone profiles from the 1% and 10% $O_2$ PAL simulations—these scenarios match the range of expected abundances for much of the Proterozoic and Phanerozoic eras, which cover more than half of Earth's history. We note that while corresponding values are taken for ozone and oxygen, we do not self-consistently model the chemical abundances of other low or trace abundance atmospheric constituents. Most of these gases are not abundant enough to impact spectra in the simulated 210–350 nm range. However, we do intend to incorporate chemically self-consistent atmospheres in future work. The planet's spectrum is modeled using the Planetary Spectrum Generator (PSG).[48,49] Using PSG, we simulate spectra of the system as viewed edge-on with the planet at quadrature, with sampling techniques and radiative transfer details given in previous work.[49]

Target SiC instrument parameters are taken from previous and current work and incorporated in PSG's instrument module. Other detector parameters are held the same for SiC and alternative detectors; this includes an assumption of detector temperatures at telescope operating temperature (not cooled from 270 K) and similar read noise and wavelength-dependent QE values for the different detectors. An NEP value of $10^{19}$ $W/\sqrt{Hz}$ ($\sim$0.00549 $e^-/sec/pixel$) for SiC is taken from earlier SiC work[3] and is scaled based upon D*/NEP values in Fig. 9, whereas an NEP value of $10^{17}$ W/$\sqrt{\text{Hz}}$ (∼55 electrons/ sec /pixel) is used as a comparison value for other detectors such as AlGaN and Si detectors.[3] This uncooled value for SiC is approximately at the required value for UV detectors listed in HWO engineering requirements,[50] and even minimal cooling would likely enable clear ability to exceed required performance. Notably, CCDs typically require cooling to −100°C to −120°C to achieve similar performance.[51–53] These values are likely overestimates for AlGaN and Si and are used to remain conservative regarding the potential improvement SiC may offer relative to other materials in the future. D* values measured for GaN and an Si Pin[3] are at least 2 and 3 orders of magnitude lower than the measured SiC in that study, and additional D* values for GaN[54–56] and Si[57,58] photodiodes are at or below those specified in that work.[3] We assume a normalized NEP value that is derived from those D* values, and then use an NEP for future GaN and Si photodiodes that improves by the same amount as SiC for our comparison, with the higher value between GaN and Si chosen for simplicity purposes. All other instrument properties (including QE, instrument temperature, and scene parameters) are kept the same.

The coronagraph model is based upon PSG's LUVOIR Extreme Coronagraph for Living Planetary Systems (ECLIPS) coronagraph model, with a modified 6.7 m inscribed diameter for the telescope (Exploratory Analytic Cases are currently examining architectures ranging from a 6 to 8 m mirror,[50] so we adopt a value within this range based upon a LUVOIR-B design[59]), a spectral resolution of $R = 7$, and an exo-zodi level of 3 zodi (3 times the level of exozodiacal dust equivalent to the zodiacal dust in the Solar System) assumed for the system. A QE value of 0.8, emissivity of 0.2, and telescope temperature of 270 K are taken from requirements for LUVOIR and are all targets for current SiC development. Figure 10 shows results for the two simulated cases (1% and 10% $O_2$ PAL) with a comparison of the performance of the different detectors. Total observation time for both cases is 3 days, which is chosen to achieve an SNR of at least 3 for the increased flux due to ozone at the longer wavelengths in the better-performing detector, which in both cases is SiC. The integration time and detector properties will vary as requirements are clarified, simulation parameters are optimized, and fabrication and detector development progress is made. However, the general relationship from the two simulations is clear and robust—the potential dark current advantages that SiC may offer from similar packaging configurations are likely to yield more robust detections of potential signatures in the NUV, by at least an order of magnitude in SNR. There are potential degeneracies between ozone and Rayleigh scattering and aerosols in this spectral region that the improved SNR offered by a superior detector could help resolve. This simulation also does not capture the advantage of visible blindness during measurement, which would lead to further improved SNR and higher reliability in measurement.

The improved noise performance that SiC may offer would translate to other potential HWO instruments as well as other observational platforms in astrophysics and other disciplines.

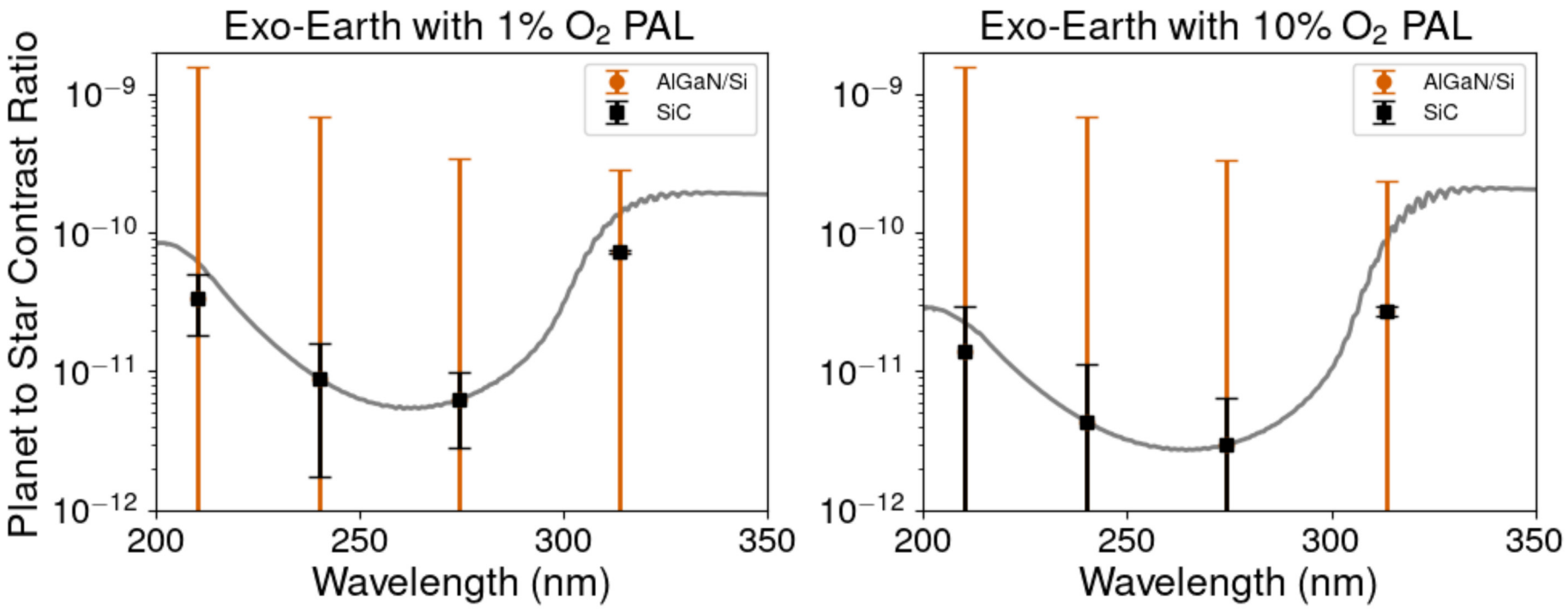

**Fig. 10** Simulated HWO spectra for an exo-Earth with atmospheric profiles corresponding to 1% and 10% $O_2$ PAL cases. We assumed an integration time of 3 days and a resolving power $R = 7$ (within the spectral range). NEP values are taken for SiC (black squares and error bars) versus other materials (orange error bars). The underlying expected contrast ratio across the Hartley-Huggins ozone band is shown in gray for reference. This figure demonstrates the unique ability of SiC to enable access to the NUV ozone signature on 3-day timescales.

Although the format and design of such detectors would vary based on the specific use cases, our team is examining potential applicability for a variety of science cases to inform additional technology and engineering advancement work on SiC detectors.

---

## Disclosures

The authors declare that there are no financial interests, commercial affiliations, or other potential conflicts of interest that could have influenced the objectivity of this research or the writing of this paper. Although the sensor and detector technologies and scientific work described in this article will be presented in proposals for future competitive calls and business opportunities, this has not influenced the dissemination of the results of the past and ongoing work in this article.

## Code and Data Availability

Data for this study are displayed in figures in the body of the paper. Underlying config files and additional data for the spectra in the HWO simulation section are available upon request. Source I-V and QE data related to NEP and $D^*$ are also available upon request.

## Acknowledgments

Our team acknowledges the following funding sources that supported the work detailed in the paper: NASA SBIR [Grant Nos. NNX12CA37C (2012), NNX15CG39P (2015), NNX17CG15C (2017), NNX17CG32P (2017), 80NSSC22PA995 (2022), 80NSSC23CA084 (2023), and 80NSSC24CA054 (2024)]. This work was also supported by the NASA Space Technology Mission Directorate Early Career Initiative Program. The work was also supported by PICASSO (Grant No. 21-PICASSO21-0016) and Ultraviolet Detector Innovation for Raman Exploration and Characterization (UV-DIRECT) of Ocean Worlds. Additional support was provided by NASA under award number 80GSFC24M0006. The team also acknowledges support through Goddard's Internal Research and Development program. P.S. would like to acknowledge support from the Goddard Space Flight Center (GSFC) Sellers Exoplanet Environments Collaboration (SEEC), which is supported by the NASA Planetary Science Division's Research Program. R.W. would like to acknowledge support from the Future Investigators in NASA Earth and Space Science and Technology program.

Biographies of the authors are not available.